\documentclass[prl,twocolumn, showpacs,nosuperscriptaddress]{revtex4}
\usepackage{amsmath}
\usepackage{amsfonts}
\usepackage{graphicx}
\usepackage{bm}

\begin{document}

\title{Inconsistency in the application of the adiabatic theorem}
\author{Karl-Peter Marzlin}
\author{Barry C.~Sanders}
\affiliation{Institute for Quantum Information Science,
University of Calgary,
2500 University Drive NW, Calgary, Alberta T2N 1N4, Canada}

\date{\today}

\begin{abstract}
The adiabatic theorem states that an initial eigenstate of a slowly
varying Hamiltonian remains close to an instantaneous eigenstate of the
Hamiltonian at a later time.
We show that a perfunctory application of this statement is problematic 
if the change in eigenstate is significant, regardless of how 
closely the evolution satisfies the requirements of the adiabatic theorem. 
We also introduce an example of a two-level system with an exactly 
solvable evolution to demonstrate the inapplicability of the 
adiabatic approximation for a particular slowly varying Hamiltonian.
\end{abstract}

\pacs{03.65.-w, 03.65.Ca, 03.65.Ta}

\maketitle

%%%%%
{\em Introduction.}---
Since the dawn of quantum mechanics \cite{ehrenfest16,born28,kato50}, 
the venerable adiabatic theorem (AT)
has underpinned research into quantum systems with adiabatically (i.e.
slowly) evolving parameters, and has applications beyond quantum physics,
for example to electromagnetic fields. The AT lays the foundation for the
Landau-Zener transition (LZT)
(including the theory of energy level crossings in
molecules) \cite{landau32}, 
for the Gell-Mann--Low theorem in quantum field theory \cite{gellmann51} on
which perturbative field theory is constructed, 
and for Berry's phase \cite{berry84}.
More recently the AT has renewed its importance in the context of
quantum control, for example concerning adiabatic passage between atomic
energy levels \cite{stirap}, as well as for 
adiabatic quantum computation \cite{adiqc}. 
Essentially the standard statement of the AT implies
that a system prepared in an instantanous eigenstate of a
time-dependent Hamiltonian will remain close to 
an instantaneous eigenstate of the
Hamiltonian provided that the Hamiltonian changes sufficiently slowly.
Here we demonstrate that the application of the standard statement of 
the AT leads to an inconsistency, regardless of how
slowly the Hamiltonian changes; although the AT itself is sound 
provided that deviations from adiabaticity are properly accounted for,
the standard statement alone does not ensure that a formal application
of it results in correct results.
In addition, we will present a simple two-level example for the failure
of the AT even in a case when all the criteria for the AT seem to be met.
The results of this paper are intended to serve as a warning that
the incautious use of the AT may produce seemingly unproblematic
results which nevertheless may be grossly wrong.

Before demonstrating the inconsistency arising from the standard statement
of the AT, it is useful to give a simple exposition of the
theorem's proof as given in Ref.~\cite{aharonov87} .
The evolution of a quantum state $|\psi(t)\rangle$
under a unitary evolution $U(t,t_0)$
is described by $| \psi (t) \rangle = U(t,t_0) |E_0(t_0) \rangle $.
We assume that the initial state $|E_0(t_0) \rangle $ is an eigenstate
of the initial Hamiltonian $H(t_0)$. The time-dependent Hamiltonian 
is related to $U(t)$ by $H(t) = \text{i} \dot{U} U^\dagger$
\footnote{We set $\hbar=1$ and $\dot{f} \equiv \text{d}f/\text{d}t$.},
such that  $| \psi (t) \rangle$ fulfills the usual 
Schr\"odinger equation.
In the instantaneous eigenbasis $\{ |E_n(t) \rangle \}$ of $H(t)$,
the state can be expressed as
$|\psi(t) \rangle = \sum_n \psi_n(t) e^{-\text{i} \int E_n}|E_n(t) \rangle $, 
where we have introduced the short-hand notation 
$\int E_n\equiv\int_{t_0}^t E_n(t^\prime )\text{d}t^\prime $.
Inserting this expansion into the Schr\"odinger equation leads
to the following differential equation for the coefficients,
\begin{equation} 
  \text{i}\dot{\psi}_n =  -\text{i} \sum_{m} e^{\text{i}\int (E_n-E_m)}\psi_m 
  \langle E_n | \dot{E}_m \rangle~.
\end{equation} 

The AT relies on the requirement that $H(t)$ is slowly varying according to
\begin{equation} 
 |\langle E_n | \dot{E}_m \rangle | \ll |E_n-E_m| \quad , \quad n\neq m~.
\label{adicrit}\end{equation} 
Transitions to other levels are then supposed to be negligible due to the
rapid oscillation arising from the phase factor
$\exp (i \int (E_n-E_m))$, yielding
\begin{eqnarray} 
  |\psi (t) \rangle  &\approx& e^{-\text{i} 
  \int E_0} e^{\text{i}\beta_0}|E_0(t)\rangle 
\label{at} \end{eqnarray} 
with $\beta_n = \text{i} \int \langle E_n | \dot{E}_n \rangle $
the geometric phase (GP) \cite{berry84}. Condition~(\ref{adicrit})
and approximation~(\ref{at}) summarize the standard statements
of the AT. 

%%%%%
{\em Proof of inconsistency.}---
The inconsistency implied by Eq.~(\ref{at}) is evident by considering the
state $|\bar{\psi} \rangle := U^\dagger(t,t_0) |E_0(t_0) \rangle $.
Using $\partial_t (U^\dagger U) =0$ it is easy to see that this
state fulfills an exact Schr\"odinger equation with Hamiltonian
$\bar{H}(t) = - U^\dagger (t,t_0) H(t) U (t,t_0)$.
To demonstrate the inconsistency, we
commence with a claim that is shown to yield a contradiction.

\noindent\textbf{Claim:} The AT~(\ref{at}) implies
\begin{equation} 
  |\bar{\psi} \rangle =   e^{\text{i} \int E_0} |E_0(t_0) \rangle~.
\label{mainres}\end{equation} 
\textbf{Proof of inconsistency:} Because $U(t_0,t_0)=1$,
result~(\ref{mainres}) fulfills the correct initial condition
so it remains to show that~(\ref{mainres}) also fulfills 
the Schr\"odinger equation:
\begin{eqnarray} 
  \text{i} \partial_t  |\bar{\psi} \rangle &=&
    - E_0(t) |\bar{\psi} \rangle
  \nonumber \\ &=&
   - E_0(t) U^\dagger U  e^{\text{i} \int E_0} |E_0(t_0) \rangle
  \nonumber \\ &\approx&
   - E_0(t)  U^\dagger e^{\text{i}\beta_0} 
    |E_0(t) \rangle
  \nonumber \\ &=&
   -  U^\dagger 
    H(t)  e^{\text{i}\beta_0}  |E_0(t) \rangle
  \nonumber \\ &\approx&
   -   U^\dagger
    H(t) U e^{\text{i} \int E_0} |E_0(t_0) \rangle
  \nonumber \\ &=&
    \bar{H}(t) | \bar{\psi} \rangle~.
\end{eqnarray} 
The AT is explicitly used in the lines with $\approx$.
However, Eq.~(\ref{mainres}) implies
\begin{eqnarray} 
 \langle E_0(t_0) | U U^\dagger |E_0(t_0) \rangle &= &
   \langle E_0(t_0) | U | \bar{\psi} \rangle
   \nonumber \\ &\approx&
   e^{\text{i} \beta_0}  \langle E_0(t_0)|E_0(t) \rangle \neq  1~,
\label{incons} \end{eqnarray} 
which is false $\Box$. 

Clearly the inconsistency is a consequence of neglecting
the deviations of Eq.~(\ref{at}) from the exact time evolution which
is free of inconsistencies. 
Stated another way, approximation (\ref{at}), without correction terms,
could only be exact in the limit of infinitesimally slow
evolution, for which the system is constant over finite time
and the evolution is indeed given by a multiplicative phase
cofactor. However, evolution is not infinitesimally slow, and
neglect of the correction terms leads to the inconsistency
demonstrated above.
To elucidate this point we define the following unitary 
transformation,
\begin{equation} 
  U_\text{AT}(t,t_0) \equiv 
  \sum_n e^{-\text{i} \int_{t_0}^t E_n} e^{\text{i}\beta_n(t)} |E_n(t) \rangle 
    \langle E_n(t_0) |\; .
\end{equation} 
The (exact) time evolution generated by $U_\text{AT}$ 
is equivalent to the standard statement (\ref{at}) of the AT 
for adiabatic motion in a finite-dimensional
Hilbert space with non-degenerate energy levels.
It is straightforward to write 
$ \bar{H}_\text{AT}(t) = -\text{i} U_\text{AT}^\dagger \dot{U}_\text{AT} $ 
in the form
\begin{eqnarray} 
 \bar{H}_\text{AT}(t)
  &=& - \sum_n E_n(t) | E_n(t_0) \rangle \langle E_n(t_0) | 
  \nonumber \\ & &
  -\text{i} \sum_{m\neq n} e^{\text{i} \int (E_n-E_m)} 
  e^{-\text{i}(\beta_n-\beta_m)}
     \nonumber
  \\ & & \times
  \langle E_n(t) |\dot{E}_m(t) \rangle \; \;
   | E_n(t_0) \rangle \langle E_m(t_0) | \; .
                    \label{alternativeDeriv}\end{eqnarray}
The second sum in this expression has the same structure
as those terms that are omitted in the adiabatic approximation,
and by omitting these terms one again arrives at the 
inconsistent result (\ref{mainres}). However,
evaluating Hamiltonian $\tilde{\bar{H}}_\text{AT}$ in 
the interaction picture with respect to the first line
of $\bar{H}_\text{AT}$, one finds
\begin{eqnarray} 
 \tilde{\bar{H}}_\text{AT}(t)
  &=&
  -\text{i} \sum_{m\neq n}  e^{-i(\beta_n-\beta_m)}
     \nonumber
  \\ & & \times
  \langle E_n(t) |\dot{E}_m(t) \rangle \; \;
   | E_n(t_0) \rangle \langle E_m(t_0) | \; .
\end{eqnarray}
In this Hamiltonian the transition matrix elements between the
different initial eigenstates
are not rapidly oscillating anymore and therefore
cannot be neglected. However, perfunctory use of the 
standard statement of the AT (\ref{at}) implicitly neglects 
such terms. 

Thus we have shown that the 
standard statement of the AT may lead to an inconsistency 
no matter how slowly the Hamiltonian is varied, but so much science rests
on the AT 
that the implications of this inconsistency are important
and require exploration. Perhaps the most important application of the AT is
the slow evolution of an initial instantaneous eigenstate $|E_0(t)\rangle$ into
a later instantaneous eigenstate $|E_0(t)\rangle$ that is meant to be quite
different; i.e.\ $\mathcal{F}_0=|\langle E_0(t)|E_0(t_0)\rangle| \ll 1$.
For example the famous LZT \cite{landau32} 
evolves a two-level molecule or atom with orthogonal
basis states $|0 \rangle $ and $|1 \rangle$ 
from $|E_0(t_0) \rangle = |0 \rangle $ 
to  $|E_0(t) \rangle = |1 \rangle $ with near-unit probability 
so that $\mathcal{F}_0 % =\langle E_0(t_0) |E_0(t) \rangle 
 \approx | \langle 0 | 1 \rangle | =0$.
On the other hand,  the quantity
$\mathcal{F}_1=| \langle E_0(0)| U U^\dagger  |E_0(0) \rangle|$ should always
be unity, but~(\ref{incons}) implies
$\mathcal{F}_1 % = | \langle E_0(0)|E_0(t) \rangle|
\approx \mathcal{F}_0 \approx 0 $.
Thus, the deviation of the overlap function 
$\mathcal{F}_0$ 
from unity is an alarm indicator for when
the AT is vulnerable to the inconsistency:
whenever $|E_0(t) \rangle$ deviates strongly
from the initial state~$|E_0(0)\rangle$, the inconsistency is a potential problem,
{\em regardless of how slowly $H(t)$ changes}. 

%%%%
{\em Counterexample of a two-level system.}---
The inconsistency introduced above 
is due to a particular inverse
time evolution which causes rapidly oscillating terms to become 
slowly varying.  One may
see this as a resonance problem which can also appear for $U$ itself.
As a specific example, consider a two-level system with
exact time evolution defined by 
\begin{equation} 
  U(t) = \exp \left ( -\text{i} \theta(t) \mathbf{n}(t)\cdot \bm{\sigma}
  \right )
  =
  \cos \theta \openone  -\text{i}  {\bf n}\cdot \bm{\sigma} \sin \theta
\label{uexample}\end{equation}   
with $\theta (t) = \omega_0 t$,
$\mathbf{n}(t) = (\cos(2\pi t/\tau ), \sin (2\pi t/\tau), 0) $
and $\bm{\sigma} = (\sigma_x, \sigma_y, \sigma_z)$ denoting the Pauli
spin vector operator.
The associated Hamiltonian can be calculated
using $H(t) = \text{i}\dot{U} U^\dagger$ and can be written in the form
$H(t) = \mathbf{R}(t) \cdot \bm{\sigma}$, with 
\begin{eqnarray} 
{\bf R}  &=& \dot{\theta} {\bf n} +  \cos \theta  \sin \theta 
   \dot{{\bf n}} + \sin^2 \theta ( {\bf n}\times \dot{{\bf n}} )
  \nonumber \\ 
  &=& \omega_0 {\bf n}(t) +  \frac{2\pi\sin (\omega_0 t)}{\tau} 
 \left ( \begin{array}{c}
  - \sin (\frac{2\pi t}{\tau} )\cos (\omega_0 t)
\\
   \cos (\frac{2\pi t}{\tau} )\cos (\omega_0 t) 
\\
   \sin (\omega_0 t)
  \end{array} \right ) 
  \nonumber \\ &\equiv&
  \omega_0 {\bf n}(t) 
  +  \frac{\sin (\omega_0 t)}{\tau} \tilde{{\bf R}}(t)\; .
\end{eqnarray} 
This Hamiltonian is similar to that of
a spin-$\tfrac{1}{2}$ system in a
magnetic field of strength proportional to $\omega_0$ that rotates 
with period $\tau$ in the $x-y$-plane. 
The exact Hamiltonian for the latter case
would correspond to $ \tilde{{\bf R}}(t)=0$; we will discuss the
importance of this difference below.
The eigenvalues of $H(t)$ 
are given by
\begin{equation} 
  E_\pm (t) = \pm |\mathbf{R}(t) | 
  = \pm\sqrt{\dot{\theta}^2 + \sin^2 \theta \dot{{\bf n}}^2 }
\end{equation} 
It is easy to show that the evolution operator 
(\ref{uexample}) fulfills requirement
(\ref{adicrit}) for adiabatic evolution as long as the vector
$\mathbf{n}$ changes slowly compared to $\omega_0$, i.e.~for 
$ \omega_0 \tau \gg 1$. The time scale $\tau$ corresponds to the
large time scale which appears in the mathematically more elaborated
forms of the AT \cite{born28,kato50,yaffe87}.
These correction terms are resonant 
\footnote{We thank M.~Oshikawa for making us aware of this.}
so that a large deviation from the AT predictions can 
accumulate over time.

To evaluate the predictions of the AT it is 
convenient to consider projection operators 
instead of the state itself.
Projectors onto eigenstates of $H(t)$, which fulfill
$H(t) |\pm(t) \rangle = \pm |{\bf R}(t)|\; |\pm(t) \rangle $,
can generally be written as
\begin{equation} 
  P_{|\pm (t)\rangle } 
  = \frac{1}{2} \left ( {\bf 1} \pm 
   \frac{{\bf R}(t) \cdot \bm{\sigma}}{|{\bf R}(t)|} \right )\; .
\end{equation} 
If we consider the evolution at time $T = \tau /2 $
and assume for simplicity that $\omega_0 T$ is a multiple
of $2\pi$, we have $ {\bf R}(T) = -{\bf R}(0)$ 
and $U(T)={\bf 1}$.
We thus find $P_{|+(T)\rangle } = P_{|-(0)\rangle }$, but 
$P_{U|+(0)\rangle } = U(T)P_{|+(0)\rangle } U(T)^\dagger = P_{|+(0)\rangle }$. 
In other words, the perfunctory prediction
\begin{equation} 
   U(T)P_{|+(0)\rangle } U(T)^\dagger \approx P_{|+(T)\rangle }
\label{adiabPredict}\end{equation} 
of the AT 
is invalid. 
Thus, whereas a resonant but weak time-dependent oscillatory term
in the evolution represents an unusual application of the AT, this
system meets the criteria of the AT and therefore casts doubt
on the general applicability of criterion (\ref{adicrit}).

%%%%%
For two-level systems, it is possible to derive a general criterion
on when the AT is bound to fail, i.e., when the quantity 
${\cal Q} := |\langle +(t) | U(t) |+(0) \rangle |^2 
= \mbox{Tr }P_{U| +(0) \rangle} P_{| +(t) \rangle}$ 
strongly deviates from one. It is evident that this criterion
depends on $U(t)$ at time $t$ only, as well as on the Hamiltonians $H(t)$ and
$H(0)$. There is no direct reference to the slow evolution
of the Hamiltonian because the criterion does not depend on $\dot{H}$.
For a unitary transformation of the form (\ref{uexample})
with general $\theta (t)$ and $\mathbf{n}(t)$ it is straightforward
to derive
\begin{equation} 
{\cal Q} = \frac{1}{2} \left (
  1 + \mathbf{n}(0) \cdot \frac{ 
  \dot{\theta} \mathbf{n} + \cos \theta \sin\theta \dot{\mathbf{n}}
  - \sin^2 \theta \mathbf{n}\times \dot{\mathbf{n}}
  }{ |\mathbf{R}|}
  \right )\; .
\label{generalCond}\end{equation} 
We have assumed that $U(0)$ is given by the identity matrix so that
$\theta(0)=0$ and $\mathbf{R}(0) = \dot{\theta}(0) \mathbf{n}(0)$. 
To examine when ${\cal Q}$ can become small we
focus on a special case of adiabatic evolutions, characterized by 
$\dot{\theta} \gg |\dot{\mathbf{n}}|$. In this case we can neglect
all terms containing $\dot{\mathbf{n}}$ such that 
$|\mathbf{R}| \approx \dot{\theta}$. We then arrive at the conclusion
that the AT is maximally violated if 
$\mathbf{n}(t)\approx - \mathbf{n}(0)$, as in the case for the example
given above. 
We remark that many other adiabatic evolutions do not fulfill
$\dot{\theta} \gg |\dot{\mathbf{n}}|$, since
it implies that $\dot{\theta}>0$ so that $U(t)$ has to become equal
to the identity again within the fast time scale $1/\dot{\theta}$.
For instance, a LZT, for which 
$\mathbf{R}(t) = \Omega \mathbf{e}_x - \dot{\Delta}_0 t/2 \mathbf{e}_z$
for constant real $\Omega$ and $\dot{\Delta}_0$, asymptotically fulfills
$\mathbf{R}(t)\approx - \mathbf{R}(-t)$, but not
$\dot{\theta} \gg |\dot{\mathbf{n}}|$ so that ${\cal Q}\approx 1$
is still valid. 

Eq.~(\ref{generalCond}) is a universal criterion for the failure
of the AT for two-level systems. Although it is likely that the
small but resonant terms in our counterexample (CE) are the cause for
this failure, a non-resonant CE to the consistency of 
the standard statement of the AT is not necessarily
excluded. This is because
Eq.~(\ref{generalCond}) depends only on the
initial and final Hamiltonian and the final unitary matrix $U$,
and therefore makes no reference to the behaviour during
the evolution.

\begin{figure}[h]
\centerline{\includegraphics[width=6cm]{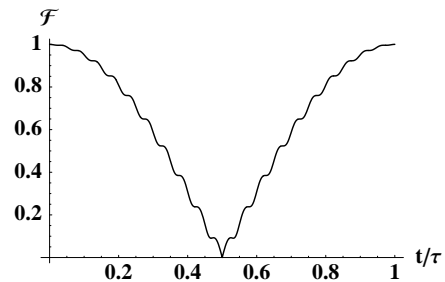}}
\caption{\label{yaffeFig} Fidelity 
${\cal F} $ between the exact evolution
(\ref{uexample}) and the instantaneous
eigenvector for $\omega_0=1$s$^{-1}$ and $\tau=2\pi\cdot 10$s$^{-1}$.}
\end{figure}

It is 
worthwhile to examine if a resonant behaviour 
as in our CE is excluded by the 
conditions imposed on the Hamiltonian in
more rigorous forms of the AT.
The two cases we do consider both require the usual gap condition 
for the energy levels, which is fulfilled in the CE. 
In addition,
Kato \cite{kato50} demands $\text{d}H(s)/\text{d}s$ to be finite for
$\tau\rightarrow\infty$, where $s=t/\tau$ is a scaled time variable.
This is the case for the CE
\footnote{Kato includes
the possibility that $H(s)$ depends $\tau$.}.
In another proof of the AT,
Avron {\em et al.} \cite{yaffe87} require the Hamiltonian
to be at least twice continuously differentiable, which is
also fulfilled by the CE
\footnote{Additional, more technical conditions 
($s$-independent closed domain, boundedness) 
are also fulfilled.}. In this case the
AT (Theorem 2.8 of Reference \cite{yaffe87})
is slightly different and states that 
$P_{U|+(0)\rangle }$ stays close to
$P_{U_A|+(0)\rangle }$, where the unitary operator $U_A(t)$ is 
generated by the modified Hamiltonian
\begin{equation} 
   H_A(t) = H(t) + i [\dot{P}_{|+(t)\rangle }, P_{|+(t)\rangle } ]
\end{equation} 
(c.f. Eq.~(1.0) and Lemma 2.2 of Ref.~\cite{yaffe87}).
For the CE presented above, we have numerically solved 
the Schr\"odinger equation (in the scaled time $s=t/\tau$) 
for the propagator $U_A$ and calculated the
fidelity (or overlap)
\begin{equation} 
{\cal F} = \mbox{Tr}\sqrt{ P_{U|+(0)\rangle }^{1/2} \;
  P_{U_A|+(0)\rangle } \;
  P_{U|+(0)\rangle }^{1/2} }
\end{equation} 
between the exact time evolution and the eigenvector subspace 
propagated with $H_A$. The result is shown in Fig.~\ref{yaffeFig}.
As in our analytical results 
the overlap becomes zero for $t/\tau = 1/2$ where the maximal
violation occurs.

Thus it
seems that the conditions on the AT are not
strict enough to exclude the CE. 
A way to exclude resonant but small behaviour may 
be to demand continuous differentiability
of $H(s)$ even in the limit $\tau \rightarrow\infty$. However,
while this would be a sufficient criterion to exclude resonances
it may not be a necessary criterion and thus could exclude other cases
in which the AT works well. Also, since it is not 
proven that resonances are the 
cause of problems, this criterion might
not exclude other cases where the AT may become problematic.

%%%%%
{\em Remarks on the validity of the AT.}---
Although
the  standard statement of the AT may be problematic in certain
applications,
previous results based on the AT are generally
not necessarily affected. The reason is that the inconsistency
is not related to the validity of the AT as an approximation
but to its  application in formal derivations.
In addition, most applications of the AT as an approximation
do not include resonant
perturbations, so that the AT should provide an excellent approximation
to the exact time evolution. This is the case, for instance, for
a real spin-$\tfrac{1}{2}$ system in a slowly rotating magnetic
field ($\tilde{{\bf R}}=0$ in the Hamiltonian above)
and for LZTs.
The correctness of the LZT may also
guarantee that the results of adiabatic quantum computation
\cite{adiqc} remain valid because, for a two-level system,
the latter can be mapped to the first.
However, if the reversed time evolution $U^\dagger(t,t_0) $ 
were to be computed using Eq.~(\ref{mainres}),
the inconsistency could yield an incorrect state. 

An example where the inconsistency associated with the AT 
poses a significant problem
is a perturbative treatment of the GP. 
For brevity we refer to Refs.~\cite{berry84,aharonov87,samuel88} 
for explanations
of the technical terms and the GP used in this paragraph.
Under the condition of parallel transport \cite{samuel88} 
the GP of an evolving state is given by the phase of
$ \langle \psi(0) | \psi(t) \rangle = \langle \psi(0) |U(t)| 
\psi(0) \rangle$. If we consider the case that the unitary
operator is slightly perturbed by an operator $P$, one can show
that for an open quantum system the associated corrections 
include terms  of the
form $\langle \psi(0) |U(t)P| \psi(0) \rangle$ and
$\langle \psi(0) |P U(t)| \psi(0) \rangle$ \cite{mgs04}.
In order to calculate these corrections one needs in particular to
find an expression for the state 
$\langle \psi(0) |U(t) = (U^\dagger (t)| \psi(0)\rangle )^\dagger $.
It is obvious that the inconsistency
would then lead to a wrong result for the GP. 

In general, a potential problem in the application of the AT could
be the presence of small fluctuations in an experiment, even if
the ideal case would not be affected by the inconsistency.
The reason is that example (\ref{uexample})
indicates that small changes can invalidate the predictions of the AT,
even if they respect the adiabadicity criterion (\ref{adicrit}).
In the two-level CE, the omission of the small terms
proportional to $\tilde{\mathbf{R}}$ in the Hamiltonian changes
a system where the AT is valid to one where it is maximally violated.
Likewise, the Hamiltonian (\ref{alternativeDeriv}) shows that it
is exactly the omission of the small terms which leads to the inconsistency.
Thus whenever adiabatic fluctuations are present in an experiment,
it seems to be necessary to check the predictions of the AT.
This could be done by checking the quantities 
${\cal F}_0$ and ${\cal Q}$ for mixed states.
To be more specific, we consider a system with fluctuations
in the classical parameters that determine its Hamiltonian. Thus,
in each run the system undergoes a unitary evolution, described
by a Hamiltonian $H^{(\alpha)} (t)$ which occurs with probability
$p_\alpha$. Assuming that the system initially is always prepared 
in an eigenstate $| E_\alpha (0) \rangle$, the density matrix for
the fluctuating system is given by
$ 
\rho(t) =\sum_\alpha p_\alpha U_\alpha (t) P_{| E_\alpha (0) \rangle} 
       U_\alpha^\dagger (t)
$. 
and one finds
\begin{eqnarray} 
  {\cal F}_0 &=& \mbox{Tr } \sum_\alpha p_\alpha  P_{| E_\alpha (0) \rangle}
    P_{| E_\alpha (t) \rangle} 
  \\
  {\cal Q} &=& \mbox{Tr }\sum_\alpha p_\alpha
    P_{ U_\alpha | E_\alpha (0) \rangle}
    P_{| E_\alpha (t) \rangle}  \; .
\end{eqnarray} 
For some index $\alpha$, the application of the AT 
may fail, but averaging over $\alpha$
could mitigate the deleterious effects.  
The exploration of the AT for fluctuating systems and mixed
states \cite{sarandy04}
is an important future direction for acertaining the validity and
limits of the AT.

In conclusion, we have demonstrated an inconsistency 
implied by the standard statement of the AT
and presented a counterexample 
of a two-level system.
Both examples  alert us to the fact that
the AT must be applied with care. 
Further work will concern testing the AT for various systems, especially
those that involve stochastic fluctations and mixed states.

Since this work first appeared as a preprint \cite{thisPreprint}, 
two subsequent preprints appeared that deal with our inconsistency.
Sarandy {\em et al.} \cite{lidar} have presented a
simplified form of the inconsistency which they regard as a validation 
of the standard statement of the AT. We interpret their work
as an alternative explanation of the cause of the inconsistency
and a second demonstration that the standard statement of the AT,
taken as it is, can lead to contradictory results. 
Pati and Rajagopal \cite{pati04} have found a different form
of inconsistency associated with the adiabatic GP.
Comments on their work and the present inconsistency have been 
made in Ref.~\cite{tong04}.
\\[3mm]
{\bf Acknowledgments}
We thank D.~Feder, S.~Ghose, D.~Hobill,
and E.~Zaremba for helpful discussions
and appreciate critical comments by 
D.~Berry, E.~Farhi, T.~Kieu, M.~Oshikawa, and A.~Pati.
We also appreciate D.A.~Lidar for informing us
about Ref.~\cite{lidar}.

%%%%

\end{document}